\documentclass[aps,prb,floatfix,twocolumn,superscriptaddress]{revtex4-1}
\usepackage[pdftex]{graphicx}
\usepackage{amsmath,amssymb,subfigure,color,verbatim}
\usepackage{yllmath}

\begin{document}
\title{Accurate Calculation of Green Functions on the $d$-dimensional Hypercubic Lattice}
\author{Yen Lee Loh}
\affiliation{Department of Physics, The Ohio State University, 191 W Woodruff Avenue, Columbus, OH  43210}
\affiliation{Department of Physics and Astrophysics, University of North Dakota, Grand Forks, ND  58202 (since Aug 2011)}
\date{2011-2-16; published 2011-6-6; arXiv version including minor corrections: 2012-10-22}
\begin{abstract}
We write the Green function of the $d$-dimensional hypercubic lattice in a piecewise form covering the entire real frequency axis.
Each piece is a single integral involving modified Bessel functions of the first and second kinds.
The smoothness of the integrand allows both real and imaginary parts of the Green function to be computed quickly and accurately for any dimension $d$ and any real frequency, and the computational time scales only linearly with $d$.
\end{abstract}
\maketitle

Lattice Green functions arise in numerous problems in combinatorics, statistical mechanics, and condensed matter physics.
In the language of condensed matter physics, the Green function $G(\rrr,\omega)$ is the propagator for a particle of energy (or frequency) $\omega$ to move a distance $\rrr$ in a tight-binding model on a lattice with a nearest-neighbor hopping.
Of particular interest is the local Green function $G(\omega) \equiv G(\rrr=\0,\omega)$, whose imaginary part is the local density of states.
Closed-form expressions exist for $G(\omega)$ on the chain, square, cubic, honeycomb, triangular, body-centred cubic (bcc), face-centred cubic, diamond, and $d$-dimensional hyper-bcc lattices in terms of elliptic integrals and hypergeometric functions (see Ref.~\onlinecite{guttmann2010} for a useful review), as well as on certain fractal lattices, \cite{schwalm1988,schwalm1992} but in general it is necessary to perform integrals or infinite sums.

%

In this work we address the problem of computing the local Green function of a $d$-dimensional hypercubic lattice with nearest-neighbour hopping $t=\half$,
	\begin{align}
	G_d (\omega) 
	&=\frac{1}{\pi^d}
		\int_0^\pi \dots \int_0^\pi 
		\frac{dk_1 \dotso dk_d}{\omega - (\cos k_1 + \dotso + \cos k_d) + i0^+}
		\label{BZIntegral}
	\end{align}
(where we have included an infinitesimal shift in the denominator as is conventional in condensed matter theory).
This function has a branch cut along the real axis for $\omega \in [-d, d]$, representing a continuous spectrum (band) of particle excitations with bandwidth $2d$.  

There are several ways of computing explicit values for $G_d (\omega)$.
The density of states of the $d$-dimensional hypercubic lattice, $A_d (\omega) = -\frac{1}{\pi} \Im G_d (\omega)$, is the convolution of $A_{d_1}$ and $A_{d_2}$ where $d=d_1+d_2$.  Since $A_d$ is known in closed form for $d=1,2,3$, this allows $A_d$ to be computed as a single convolution for $d=4,5,6$, but for $d\geq 7$ two or more nested integrations are required.  
It is simple to derive the Laurent series for $G_d(\omega)$ from a multinomial expansion,\cite{guseinov2007,mamedov2008} 
but due to the branch points at $\pm d$, the Laurent series only converges for $\left| \omega \right| \geq d$, on or outside a circle of radius $d$ (barring techniques such as Borel summation).
$G_d(\omega)$ may be written as a Fourier integral involving Bessel functions (in which case the accuracy of numerical integration is limited by oscillations of the integrand), \cite{maradudin1960,oitmaa1971}
or as an Laplace transform integral involving the modified Bessel function $I_0$ (which only converges for frequencies outside the band edge).\cite{maradudin1960}      
This work presents integral representations for both real and imaginary parts of $G_d (\omega)$ applicable on the entire real axis, both inside and outside the band; the integrands are smooth and well-behaved, which makes it easy to obtain numerical results to high precision.

	\begin{figure}[!htb]
	\subfigure[~$G_d (\omega)$]{
		\includegraphics[width=0.45\columnwidth]{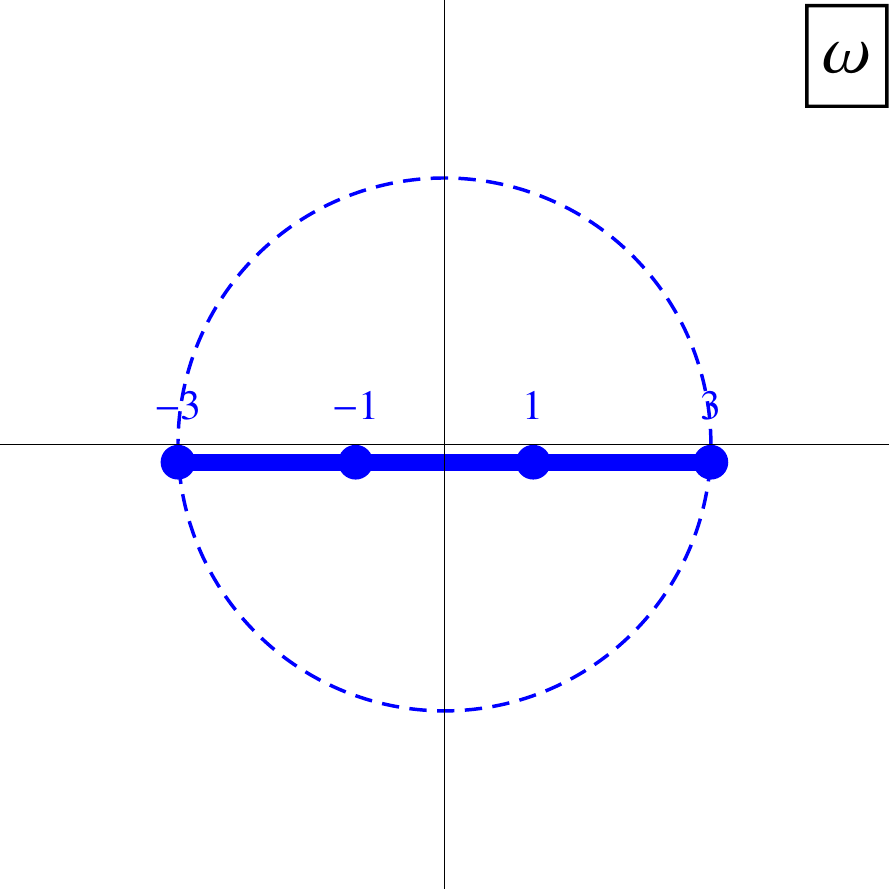}
	}
	\subfigure[~$e^{i\omega t} H(t)^{d-n} h(t)^n$]{
		\includegraphics[width=0.45\columnwidth]{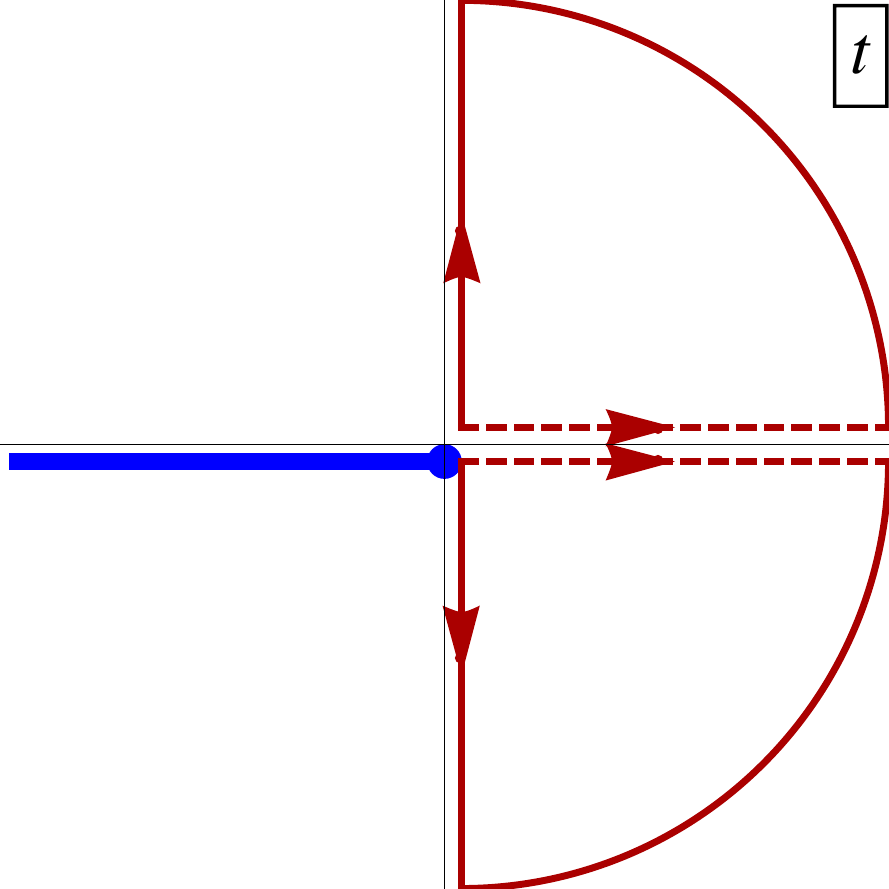}
	}
	\caption{
	\label{Argand}
	(a) Structure of the Green function in the complex plane, illustrated for $d=3$.
	Thick lines represent branch cuts and dots represent branch points.
	The Laurent series converges only on and outside the dashed circle.
	In contrast, the method of this paper applies for any $\omega$ on the real axis.
	(b) Structure of Eq.~\eqref{Equation4} in the complex plane.
	The branch cut runs below the negative real axis,
	terminating in a logarithmic branch point.
	The integration path runs along the positive real axis,
	and is deformed to run along the positive or negative imaginary axis
	depending on the value of $\omega+d-2n$.
	}
	\end{figure}

Equation \eqref{BZIntegral} gives the Green function as a multidimensional integral.
Fourier-transforming to the time-domain, performing the wavevector integrals, and Fourier-transforming back to the frequency domain leads naturally to a single-integral formula in terms of Bessel $J$ functions:
	\begin{align}
	G_d (t)
	&=\int_{-\infty}^\infty \frac{d\omega}{2\pi}~ e^{-i\omega t} G_d (\omega)
\nonumber\\
	&=-i\int_0^\pi \frac{dk_1 \dotso dk_d}{\pi^d}		~
		\Theta(t) e^{ -it(\cos k_1 + \dotso + \cos k_d) }
\nonumber\\
	&=-i\Theta(t) 	J_0(t) ^d ,
\nonumber\\
	G_d (\omega) 
	&=-i \int_0^\infty dt~ e^{i\omega t} J_0(t) ^d
	\elabel{BesselFormula}
		.
	\end{align}
From \eref{BesselFormula} one can quickly estimate $G_d (\omega)$ on the real axis using ordinary quadrature or fast Fourier transform methods.
\cite{maradudin1960,oitmaa1971}
However, the oscillatory nature of the integrand limits the accuracy of this approach to a few digits.
In situations like this, a useful strategy is to choose an integration path within the complex plane along which the integrand is smooth and non-oscillating.
\cite{maradudin1960}
However, for frequencies within the band ($\left| \omega \right| < d$) this trick is not directly applicable because the integrand grows exponentially in both the upper half-plane (UHP) and lower half-plane (LHP).  Therefore, we split the Bessel function into Hankel functions $H(t) \equiv H_0^{(1)} (t)$ and $h(t) \equiv H_0^{(2)} (t)$,
	\begin{align}
	G_d (\omega) 
	&=-i \int_0^\infty dt~ e^{i\omega t} \left[ \frac{H(t) + h(t)}{2} \right] ^d
	,
	\end{align}
and perform a binomial expansion to obtain
	\begin{align}
	G_d (\omega) 
	&=-\frac{i}{2^d}
		\sum_{n=0}^d \tbinom{d}{n} 
		\int_0^\infty dt~ e^{i\omega t} 
		H(t)^{d-n} h(t)^{n}  
		.
	\label{Equation4}
	\end{align}
The asymptotic behavior of the integrand, ignoring power-law factors, is
$e^{it(\omega + d - 2n)}$.  
If $\omega + d - 2n > 0$, the integrand decays exponentially into the UHP, so Jordan's lemma allows us to deform the integration contour into the positive imaginary axis.  Conversely, if $\omega + d - 2n < 0$, we can deform the contour into the negative imaginary axis.  Thus,
	\begin{align}
	G_d (\omega) 
	&=-\frac{i}{2^d}
		\sum_{n=0}^d \tbinom{d}{n}
		~\bigg[
	\nonumber\\&{}
			~~~
			\Theta(\omega + d - 2n) ~
			\int_0^{i\infty} dt~ e^{i\omega t} H(t)^{d-n} h(t)^{n}
	\nonumber\\&{}
			+
			\Theta(-(\omega + d - 2n))
			\int_0^{-i\infty} dt~ e^{i\omega t} H(t)^{d-n} h(t)^{n}
		\bigg]
		.
		\nonumber
	\end{align}
Substituting $t=i\tau$ and $t=-i\tau$ respectively into the two integrals collapses both integrations onto $\tau \in (0,\infty)$, allowing us to combine integrands:
	\begin{align}
	G_d (\omega) 
	&=
		\frac{1}{2^d} \int_0^\infty d\tau~ 
		\sum_{n=0}^d \tbinom{d}{n}
		~\bigg[
	\nonumber\\&{}
			~~~
			\Theta(\omega + d - 2n) ~
			e^{-\omega\tau} 
			H(i\tau)^{d-n} h(i\tau)^{n}  
	\nonumber\\&{}
			-
			\Theta(-(\omega + d - 2n))
			e^{\omega\tau} 
			H(-i\tau)^{d-n} h(-i\tau)^{n}  
		\bigg]
		.
		\nonumber
	\end{align}
The sum can be split into two terms, one for $n<j$ and one for $n\geq j$, where $j(\omega)=\lfloor \frac{\omega+d}{2} \rfloor$ is a staircase function:
	\begin{align}
	G_d (\omega) 
	&=
		\frac{1}{2^d}
		\int_0^\infty d\tau~ 
		\bigg[
		e^{-\omega\tau} 
		\sum_{n=0}^{j} 
			\tbinom{d}{n}
			H(i\tau)^{d-n} h(i\tau)^{n}  
	\nonumber\\&{}
	~~~~~~~~~~~~
		-
		e^{\omega\tau} 
		\sum_{n=j+1}^{d}
			\tbinom{d}{n}
			H(-i\tau)^{d-n} h(-i\tau)^{n}
		\bigg]
		.
	\elabel{intrep2}
	\end{align}
The integrand in \eref{intrep2} has integrable logarithmic singularities at $\tau=0$.  When $\omega \in \{-d,-d+2,\dotsc,d\}$, corresponding to van Hove singularities, the integrand exhibits power-law decay at large $\tau$; otherwise, it decays exponentially.  These asymptotic behaviors are easily dealt with using standard transformations such as those built into \emph{Mathematica}'s integration routines.  At intermediate $\tau$ the integrand is a smooth function that can be integrated quickly and accurately.


	\begin{figure*}[!htb]
	\includegraphics[width=0.48\textwidth]{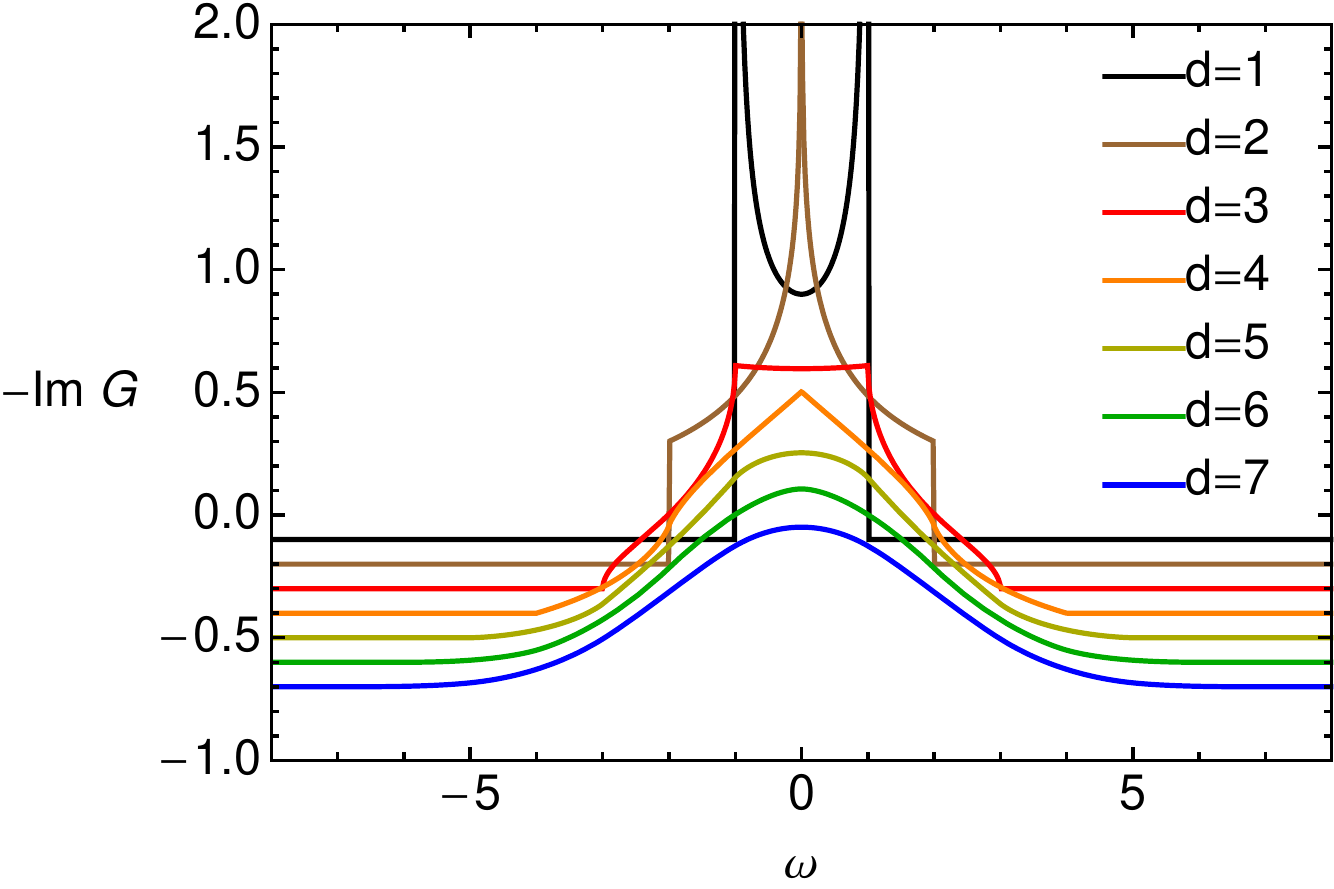}
	\includegraphics[width=0.48\textwidth]{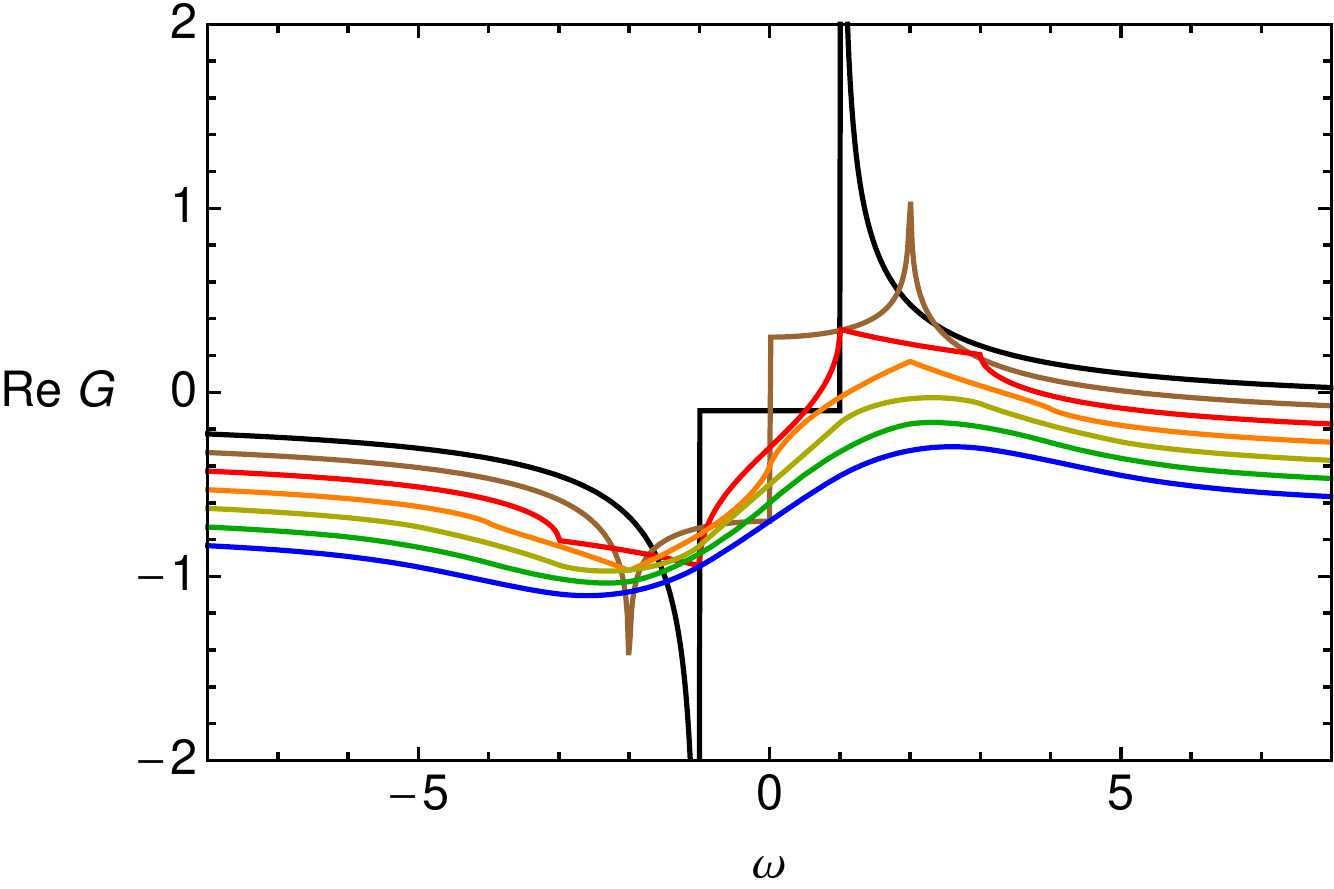}
	\caption{
	\label{GreenFunctions}
	Real and imaginary parts of hypercubic lattice Green functions
	for $d=1,2,3,4,5,6,7$ (shifted by $-0.1d$ for clarity).
	The van Hove singularities are almost unnoticeable for $d\geq 5$.
	As $d\rightarrow \infty$, $\Im G(\omega)$ tends to a Gaussian as expected.
	}
	\end{figure*}

For computational purposes, we can further optimize \eref{intrep2} by explicitly writing out the real and imaginary parts of the Hankel functions.  For $\tau>0$, 
	\begin{align}
	H(i\tau) &= -i K(\tau)    ,\nonumber\\
	H(-i\tau) &= -i K(\tau) + I(\tau) ,\nonumber\\
	h(i\tau) &=  i K(\tau) + I(\tau) ,\nonumber\\
	h(-i\tau) &=  i K(\tau)     ,\nonumber
	\end{align}
where for convenience we have defined
$	K(\tau) \equiv \tfrac{2}{\pi} K_0 (\tau) $ and
$ I(\tau) \equiv 2 I_0 (\tau) $
where $K_0$ and $I_0$ are modified Bessel functions (as implemented in \emph{Mathematica}).
Substituting in these relations, performing a binomial expansion, and rearranging summations leads to an integral containing a sum of $d+1$ products of modified Bessel functions,
	\begin{align}
	G_d (\omega) 
	&=
		\frac{1}{2^d}
		\int_0^\infty d\tau~ 
		~\bigg[
		e^{-\omega\tau} 
		\sum_{m=0}^{j} 
			C_{jm}
			K^{d-m}(\tau)
			I^{m} (\tau)
	\nonumber\\&{}
	~~~~~~~~~~~~~~~~~~~
		-
		e^{\omega\tau} 
		\sum_{m=0}^{d-j-1} 
			D_{jm}
			K^{d-m} (\tau)
			I^{m} (\tau)
		\bigg]
	,
	\end{align}
where the coefficients
	\begin{align}
	C_{jm}
	&=
		\sum_{n=m}^{j} 
			\tbinom{d}{n}
			\tbinom{n}{m}
			i^{2n - d - m}
	\nonumber\\
	D_{jm}
	&=\sum_{n=m}^{d-j-1} 
			\tbinom{d}{n}
			\tbinom{n}{m}
			i^{d + m - 2n}
	\end{align}
can be precomputed.  It should be noted that the above sums can be evaluated in closed form using the properties of binomial coefficients.  
As an illustration, setting $d=3$ leads to an integral formula for the cubic lattice Green function,
\begin{widetext}
	\begin{align}
	G_3 (\omega) 
	&= 
		\frac{1}{8}		\int_0^\infty d\tau~ 
	\begin{cases}
	-I^3 e^{\omega\tau} 	                                                       & \omega \leq -3 \\
	-3 I K^2 e^{\omega\tau} + iK(2 K^2 \cosh \omega\tau - 3I^2 e^{\omega\tau})  & -3 \leq \omega \leq -1 \\
	6 I K^2 \sinh \omega\tau - 4i K^3 \cosh \omega\tau                          & -1 \leq \omega \leq 1 \\
	3 I K^2 e^{-\omega\tau} + iK(2 K^2 \cosh \omega\tau - 3I^2 e^{-\omega\tau}) & 1 \leq \omega \leq 3 \\
	I^3 e^{-\omega\tau} 	                                                       & \omega \geq 3 
	\end{cases}
	\elabel{CubicLatticeIntegralFormula}
	\end{align}
\end{widetext}
(where $\tau$ arguments of $K$ and $I$ have been suppressed for clarity).
\eref{CubicLatticeIntegralFormula} agrees with existing formulas for $\left| \omega \right| \geq 3$, \cite{maradudin1960}
and it can be implemented with good accuracy.  For example, $G_3(0) = -0.8964407887768\dots$, which agrees with the closed-form result \cite{joyce1972} to 12 digits.
The Green functions for $d=1$ up to $d=7$ are plotted in Fig.~\ref{GreenFunctions}.

An elegant feature of this method is that it gives a piecewise representation of the Green function, where van Hove singularities emerge naturally from sudden changes in the integration path.  This is in contrast with Laurent, Fourier, and Chebyshev series methods, where the summation proceeds blindly and requires a large number of terms to approximate the singular behavior (if it converges at all).

The approach can easily be generalized to off-diagonal Green functions $G(\rrr,\omega)$.  However, because it relies on the factorization property embodied in \eref{BesselFormula}, there is no obvious way to generalize it to non-hypercubic lattices.  

Calculating properties of lattices in $d>3$ dimensions is admittedly an academic pursuit with rather contrived applications (for example, $G_6(\omega)$ is the local two-particle spectral function of a cubic lattice).
Nevertheless, the work presented here is valuable as a pedagogical example as how complex analysis can vastly improve the accuracy of numerical integration.  The insights obtained from this exercise may well be useful in more practical areas.



\bibliography{hlgf}
\end{document}